\documentclass[aps,prl,amsmath,amssymb,reprint,preprintnumbers,superscriptaddress,longbibliography,nobalancelastpage,nofootinbib]{revtex4-2}

\usepackage{xcolor}
\usepackage{accents}
\usepackage[math]{cellspace}
\usepackage{mathtools}
\usepackage{hyperref}
\usepackage{multirow}
\usepackage{autofigs}

\graphicspath{{figures/}}

\begin{document}

\preprint{FERMILAB-PUB-20-558-E}

\title{Interferometric Constraints on Spacelike Coherent Rotational Fluctuations}

\author{Jonathan~W.~Richardson}
%\email{jwr@caltech.edu}
\affiliation{California Institute of Technology, Pasadena, CA}

\author{Ohkyung~Kwon}
\email{kwon@uchicago.edu}
\affiliation{University of Chicago, Chicago, IL}
\affiliation{Korea Advanced Institute of Science and Technology, Daejeon, Republic of Korea}

\author{H.~Richard~Gustafson}
%\email{gustafso@umich.edu}
\affiliation{University of Michigan, Ann Arbor, MI}

\author{Craig~Hogan}
%\email{craighogan@uchicago.edu}
\affiliation{University of Chicago, Chicago, IL}
\affiliation{Fermi National Accelerator Laboratory, Batavia, IL}

\author{Brittany~L.~Kamai}
%\email{bkamai@ligo.caltech.edu}
\affiliation{California Institute of Technology, Pasadena, CA}
\affiliation{University of California Santa Cruz, Santa Cruz, CA}

%\author{Robert~K.~Lanza}
%\email{bobbylanza@gmail.com}
%\affiliation{Massachusetts Institute of Technology, Cambridge, MA}

\author{Lee~P.~McCuller}
%\email{lee.mcculler@gmail.com}
\affiliation{Massachusetts Institute of Technology, Cambridge, MA}

\author{Stephan~S.~Meyer}
%\email{meyer@uchicago.edu}
\affiliation{University of Chicago, Chicago, IL}

\author{Chris~Stoughton}
%\email{stoughto@fnal.gov}
\affiliation{Fermi National Accelerator Laboratory, Batavia, IL}

\author{Raymond~E.~Tomlin}
%\email{tomlin@fnal.gov}
\affiliation{Fermi National Accelerator Laboratory, Batavia, IL}

\author{Rainer~Weiss}
%\email{weiss@ligo.mit.edu}
\affiliation{Massachusetts Institute of Technology, Cambridge, MA}

%\collaboration{Holometer Collaboration}

\begin{abstract}
Precision measurements are reported of the cross-spectrum of rotationally-induced differential position displacements in a pair of colocated 39~m long, high power Michelson interferometers. One arm of each interferometer is bent $90^{\circ}$ near its midpoint to obtain sensitivity to rotations about an axis normal to the plane of the instrument. The instrument achieves quantum-limited sensing of spatially-correlated signals in a broad frequency band extending beyond the 3.9~MHz inverse light travel time of the apparatus. For stationary signals with bandwidth $\Delta f > 10\;\mathrm{kHz}$, the sensitivity to rotation-induced strain $h$ of classical or exotic origin surpasses $\mathrm{CSD}_{\delta h} < t_P / 2$, where $t_P = 5.39 \times 10^{-44}\;\mathrm{s}$ is the Planck time. This measurement is used to constrain a semiclassical model of nonlocally coherent rotational degrees of freedom of spacetime, which have been conjectured to emerge in holographic quantum geometry but are not present in a classical metric.
\end{abstract}

%%% Research Areas > Gravitation > Experimental studies of gravity
%%% Techniques > Experimental Techniques > Gravitational wave detectors
%\keywords{}

\maketitle

%--------------------------------------------

\vspace*{-12.5pt}In this Letter, we report the results of an interferometric experiment designed to measure spatially coherent rotational fluctuations of a macroscopic system, on time scales faster than its light crossing time. The instrument, a reconfiguration of the Fermilab Holometer~\cite{chou2017}, consists of two colocated and coaligned $L=38.9\;\mathrm{m}$ long power-recycled Michelson interferometers, each operating at 1.3~kW power with a mean shot-noise-limited sensitivity of $2.7 \times 10^{-18}\;\mathrm{m/\sqrt{Hz}}$. The Holometer programme is designed with two measurement configurations which collectively constrain a wide class of possible coherent spacelike fluctuations within a plane. For the first experiment~\cite{chou2016,chou2017a,chou2017b}, the light paths were entirely colinear and extended radially in orthogonal directions, as in gravitational wave detectors. In the present study, a new geometry is probed where one arm of each interferometer is bent $90^{\circ}$ near its midpoint to obtain an instrumental response to rotational modulations: rotations about an axis normal to the interferometer plane couple to the sensed degree of freedom, the differential arm length (DARM), $\delta l$, at high frequencies. Rotationally-induced displacement would not have been detected in previous experiments.

The DARM signals of the two interferometers are sampled at 50~MHz. Each achieves shot noise-limited displacement sensitivity in a broad frequency band from 1.1~MHz to 20~MHz, extending beyond the 3.85~MHz inverse light travel time of the apparatus. The two DARM signals are cross-correlated, averaging down below shot noise to a sensitivity of $6.1 \times 10^{-21}\;\mathrm{m/\sqrt{Hz}}$ at 9.92~kHz resolution to stationary signals common to both interferometers. For each frequency bin, we average over $3.9 \times 10^{10}$ independent spectral measurements. In units of strain, $\delta h = \delta l / L$, our sensitivity to rotationally-induced displacement noise surpasses a milestone $\mathrm{CSD}_{\delta h} < t_P / 2$, where $t_P \equiv \sqrt{\hbar\hspace{.07em}G/c^5} = 5.39 \times 10^{-44}\;\mathrm{s}$ is the Planck time.\vspace{-3.1pt}\newpage

\vspace*{-11.2pt}Our measurements are sensitive to a broad variety of coherent phenomena. In the context of a classical spacetime, they can constrain models of exotic new physical fields, including, for example, axion-like dark matter with a coherence scale comparable to or exceeding the apparatus size. Depending on the coupling mechanisms of the new fields, a cross-spectral signal can arise from a variety of correlated physical effects in the two interferometers~\cite{Grote:2019uvn}. Probing for possible phenomena beyond the framework of local field theory, our measurements are used here to constrain a model of coherent, spacelike rotational fluctuations of spacetime, not described by local fluctuations of a classical metric, which can arise in a holographic quantum geometry~\cite{Hogan:2015b,Hogan:2016}.

It is not known how the spacetime locality built into classical relativity can be reconciled with quantum mechanics; the active gravity from the mass-energy of a nonlocal physical state leads to a breakdown of universal causal consistency~\cite{Zych_2019}. Predictions for physical correlations at this intersection vary widely~\cite{CohenKaplanNelson1999,*cohen2021gravitational,Banks:2018ypk,*Banks:2020dus,Hooft2018,Giddings:2019vvj,Verlinde2019,Verlinde:2019ade,Parikh:2020nrd,*Parikh:2020kfh,Carney_2019,Howl_2021} and depend critically on the nature of macroscopic quantum coherence in geometrical states. For example, if classical spacetime emerges from thermodynamics of covariant causal structures~\cite{Jacobson1995,Jacobson:2015hqa}, it is possible that all null surfaces, such as black hole horizons and light cones, represent nonlocally coherent quantum objects at all scales.

One estimate of spacetime fluctuations from holography~\cite{Verlinde2019,Verlinde:2019ade,Hogan:2015b,Hogan:2016}, analogous to the standard quantum positional uncertainty $\langle\hspace{.07em}\Delta x^2\hspace{.07em}\rangle > \hbar\hspace{.07em}\tau/m$ of a mass~$m$ over time~$\tau$, is that a causal diamond surface of radius $R= c\hspace{.07em}\tau$ fluctuates coherently with a variance $\langle (\delta R\hspace{.07em}/R)^2\hspace{.07em}\rangle \sim t_P/\tau$, or a strain power spectral density of $t_P\;(\hspace{.07em}\sim 10^{-44}\,{\rm Hz}^{-1})$ over a bandwidth $1/\tau$. This corresponds to the scale of coherent quadrupolar distortions a black hole horizon needs in order to radiate at the standard Hawking flux, one graviton of wavelength~$c\hspace{.07em}\tau$ per time~$\tau$. This estimate is controversial because, due to the coherence on null structures, even Planck-scale uncertainties can create spacelike fluctuations of causal surfaces much larger than expected in standard local effective field theory and linearized gravity, by a factor $\sim \tau/\hspace{.07em}t_P$.

Here, we experimentally test one such model~\cite{Hogan:2016} in which correlated interferometer signal fluctuations arise from rotational uncertainties in the laboratory inertial frame as local spacetime is emergently defined from a quantum system without a background. The modeled instrument response to rotational uncertainty scales with a parameter normalizing the displacement spectral density, nominally $\eta \approx t_P$, connected to information limits on timelike world lines and their causal diamond boundaries. We constrain this parameter to $\eta < 0.25\,t_P$ for models with one rotational axis; values much less than $t_P$ correspond to an information excess in violation of the holographic entropy bound. Our result still leaves untested general 3D models with two incompatible rotational observables, targeted by future experiments~\cite{grote2020}.

%--------------------------------------------
\paragraph{\textbf{Experimental design.}}
The current bent-arm interferometers are a reconfiguration of the original instrument described in detail by Ref.~\cite{chou2017}. The vacuum infrastructure, sensing and control system, and almost all optical components are carried over. Here, we present an overview of the detector design with an emphasis on the changes made for rotational sensitivity.

\begin{figure}[t]
  \centering
  \includegraphics[width=1.\columnwidth,trim=0.02in 0.19in 0.06in 0.07in]{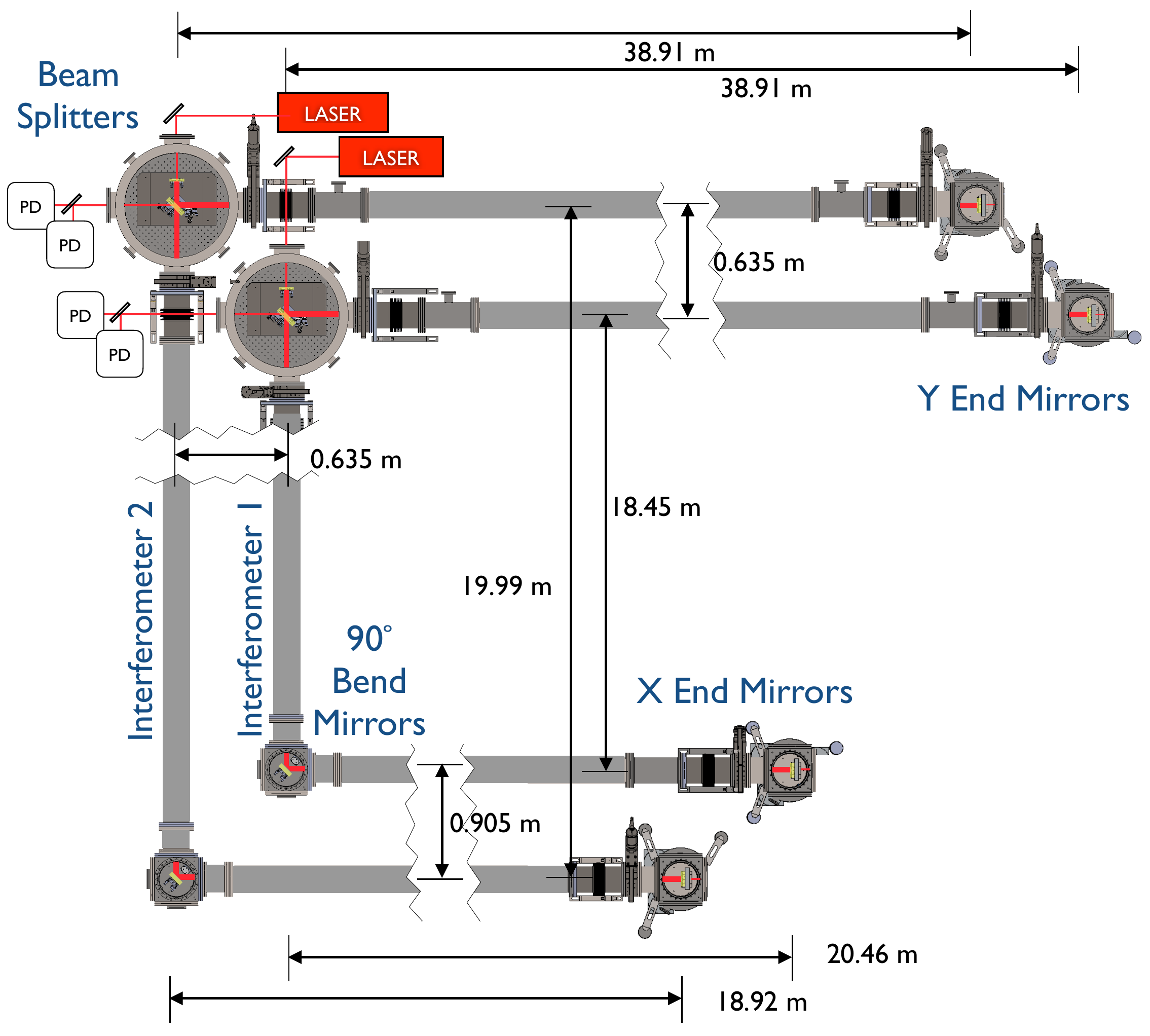}
  \caption{Layout of the dual 40~m interferometer system. In this experiment, the X-arm of each interferometer is bent $90^{\circ}$ near its midpoint. Dimensions are as marked, with the horizontal plane of interferometer~1 sitting 0.15~m higher than that of interferometer~2. The vacuum chambers are rendered with transparent tops; red lines indicate the positions of the laser beams, with heavier lines indicating the power-recycled cavity. Two radiofrequency photoreceivers sense the output intensity of each interferometer. A diagram of the post-detection signal flow is shown in Fig.~1 of Ref.~\cite{chou2016}.\vspace{-0.08in}}
  \label{fig:layout}
\end{figure}

The Holometer consists of two 38.9~m power recycled Michelson interferometers, separated by 0.9~m beamsplitter-to-beamsplitter and coaligned. Fig.~\ref{fig:layout} displays their layout. In each interferometer, continuous-wave $\lambda=1064$~nm laser light is injected to a beamsplitter and routed along two distinct optical paths to distant end mirrors, where it is retroreflected. The returning beams coherently interfere at the beamsplitter with an output intensity sinusoidally dependent on $\delta l$.

To produce a linear optical response to small DARM fluctuations, each interferometer is operated at an offset of approximately 1~nm from a dark fringe. A digital servo senses fluctuations in the power exiting the antisymmetric port and feeds back differential signals to piezoelectrically actuated end mirrors at frequencies up to 600~Hz~\cite{chou2017}. It maintains 50~pm RMS residual motion, as measured through the 16~kHz Nyquist frequency of the control system. The remaining power exiting the symmetric port is reflected back into the interferometer using a $T=1000\;\mathrm{ppm}$ transmission mirror. This mirror forms an overcoupled Fabry-Perot cavity with the interferometer whose 3.85~MHz free spectral range is determined by the average arm length. The input laser frequency is locked to the average arm length via the Pound-Drever-Hall technique~\cite{Black2001}, achieving a typical resonating power of 1.3~kW from 1.0~W of injected power. A separate radiofrequency (RF) data acquisition system samples the interferometer output intensities at 50~MHz and computes their cross-spectral density in real time.

When the light propagates along entirely colinear paths extending radially, as in the original instrument, rotations about an axis normal to the interferometer plane at the beamsplitter---the point of measurement---displace the end mirrors transversely to the optical axis, inducing no change in arm length. To obtain a first-order coupling of rotation to DARM, one arm of each interferometer is bent $90^{\circ}$ near its midpoint, as illustrated in Fig.~\ref{fig:layout}. In this configuration, such rotations displace both mirrors in each X-arm along its optical axis, resulting in a net arm length modulation at high frequencies. Although layouts more optimal for rotational sensitivity are possible, our design was chosen as a practical balance between infrastructure constraints, maximizing the reusability of the previous system, and the required integration time for a statistically-conclusive model test. The reconfigured instrument also remains sensitive to purely translational, non-rotational displacements, but such sources were excluded to similar precision by the previous Holometer experiment~\cite{chou2016,chou2017a} and their coupling strength here is smaller. Measured limits on the possible couplings of environmental RF noise are presented in the \textit{Supplemental~Material}. Throughout, we quote our measurements in units of the {\it effective} linear arm length displacement, rather than in angular units, because the angular calibration is dependent on the model of rotationally-induced displacement. For example, under the model tested here, the rotations are non-rigid-body in nature~\cite{Hogan:2016}. The arm length displacement can also be readily converted to units of differential light phase via the multiplicative factor $4 \pi / \lambda$.

Each $90^{\circ}$ arm bend is implemented through the addition of a 2~inch diameter, 1/2~inch thick highly-reflective mirror. The mirrors are fabricated from Corning 7980 0A low-inclusion fused silica substrates and polished to sub-nm flatness, with a 5~arc-minute wedge. They are coated for high $p$-polarization reflectivity at $45^{\circ}$ incidence and have a measured transmission of $T < 10\;\mathrm{ppm}$. The interferometers do not have output mode cleaners, so contrast defect reduces their sensitivity. To minimize the defect, the same coating is applied to both sides, mitigating distortions of the beam wavefront induced by mechanical stress in the coating. Each bend mirror is mounted in a Newport 8822-UHV two-axis picomotor mirror mount placed on a two-stage seismic isolation platform. The entire assembly is housed inside a 10~inch six-port vacuum cube. Similarly to the end mirror stations~\cite{chou2017}, high-frequency vibration reduction is achieved with a passive system of masses mounted on three 19~mm diameter Viton balls, here with two stacked stages using 10.5~kg steel masses. Each vacuum cube is mounted on a steel slab set on concrete footings originally used by the vacuum tubes of the straight-arm configuration. Along the new section of arm, the vacuum tubes are supported by new 0.3~m diameter concrete pillars set 1.9~m deep. The arm tubes are mechanically decoupled from the corner mirror stations using hydroformed stainless steel bellows. The two relocated end stations are mounted on separate steel plates, each weighing 2300~kg and resting on three 0.3~m diameter concrete pillars set 1.9~m deep.

%--------------------------------------------
\paragraph{\textbf{Measurement.}}
The differential arm length signals of the two interferometers are sensed by photoreceivers at each antisymmetric port and synchronously digitized at 50~MHz. As in previous studies~\cite{chou2016,chou2017a,chou2017b}, the time series are characterized by their cross-spectral density (CSD), a measure of the correlated interferometer noise power,%\vspace{1pt}
\begin{equation}\begin{split}
    \mathrm{CSD}&\left[\delta l_1, \delta l_2\right](f) \\
    &\!\!\!\equiv\int_{-\infty}^{\infty} \langle \hspace{.07em} \delta l_1(t) \,\hspace{.11em} \delta l_2(t - \tau) \hspace{.07em}\rangle_{t} \,\hspace{.11em} e^{- 2\pi i \tau f} \, d\tau \;,%\vspace{1pt}
\label{CSD}
\end{split}\end{equation}
where the subscripts denote interferometers 1 and 2 (see Fig.~\ref{fig:layout}), $f$ is frequency, and $\langle\,\rangle_t$ represents a time average over extended signal streams. Eq.~\ref{CSD} can be equivalently expressed in units of strain as $\mathrm{CSD}_{\delta h} = \mathrm{CSD} / L^2$, which has the dimensionality of inverse frequency, or time. For frequencies in the 100~kHz to 20~MHz band, the interferometer signals are calibrated to absolute length to better than 1\% statistical uncertainty with 5-7\% systematic uncertainty, limited by the linearity of the photoreceivers. The spectral densities are estimated via Welch's periodogram method~\cite{Welch:1967} using a Hann window, 50\% overlapped segments, and a discrete Fourier transform size of $N_\mathrm{DFT}=2^{16}$. The resulting 763~Hz wide frequency bins are subsequently rebinned to 9.92~kHz resolution via frequency-space averaging, accounting for the bin-to-bin covariance due to spectral leakage. Each 9.92~kHz frequency bin thus constitutes an independent measurement. Details about the data pipeline and signal calibration are described in \S5 and \S7 of Ref.~\cite{chou2017}, respectively.

\begin{figure}[t]
  \centering
  \includegraphics[width=1.\columnwidth,trim=0.09in 0.21in 0.07in 0.00in]{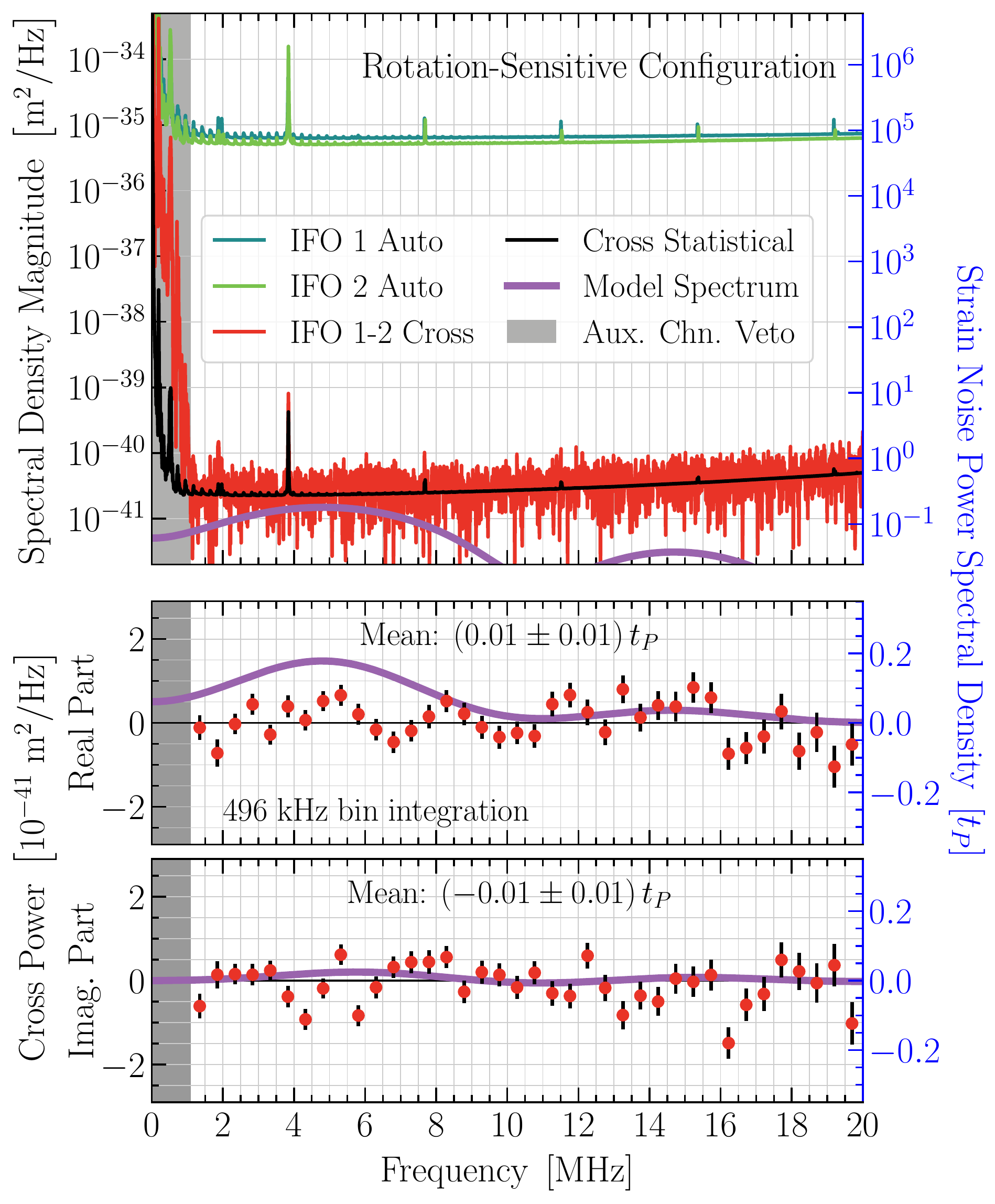}
  \caption{Measured power and cross spectral densities of the interferometer signals, shown in displacement units (left axis) and in strain units normalized to Planck time (right axis). The spectra are averaged over 1098~hours of dual interferometer time series data, or $3.9 \times 10^{10}$ independent spectral measurements per 9.92~kHz bin. Top panel: Spectra magnitudes at 9.92\,kHz resolution. The teal and green curves show the power spectral density of each interferometer (IFO). The red curve shows the cross spectral density (CSD), whose statistical sensitivity is indicated by the black trace. Bottom panels: Real and imaginary CSD components, rebinned to 496~kHz resolution. Error bars reflect the combined $1\sigma$ statistical uncertainty and 10\% systematic calibration uncertainty~\cite{chou2017}. For reference, all panels are overlaid with a semiclassical model spectrum of quantum geometrical fluctuations (purple curves) reproduced from Ref.~\cite{Hogan:2016}, but adjusted for the actual dimensions of the as-built instrument. Bins showing excess coherence with environmental channels are vetoed (grey shading).\vspace{-0.08in}}
  \label{fig:main_Result_v2}
\end{figure}

Figure~\ref{fig:main_Result_v2} shows the spectral averages of 1098~hours of dual interferometer time series data, collected in five observing runs from April~2017 to August~2019. All runs are found to be statistically consistent with one another, and their chronology is detailed in the \textit{Supplemental~Material}. The power spectral densities, or auto spectra, of the individual interferometer signals (teal and green curves, top panel) are dominated by laser amplitude and phase noise and seismic/mechanical noise below 1.1~MHz and by photon shot noise at higher frequencies. The lines spaced every 3.85~MHz are transmission resonances of the power-recycling cavity (free spectral ranges), at which frequencies input laser noise sidebands transmit to the antisymmetric port unattenuated. The weaker set of features spaced every 225~kHz in the auto spectra are thermally-excited bulk modes of the fused silica substrates of the end mirrors, bend mirror, and beamsplitter.

In the CSD of the two interferometer signals (red curves or data points, all panels), these uncorrelated sources of noise are averaged down over $3.9 \times 10^{10}$ independent spectral measurements per 9.92~kHz bin to attain a sensitivity more than five orders of magnitude below the quantum sensing limit of either instrument individually. Overall, the noise level is flat, but rises slightly at higher frequencies due to phase noise between the two independent sample clocks, which leads to a growing decoherence between the digitized interferometer signals with higher frequency~\cite{chou2017}. The excess coherence in noise below 1.1~MHz is due to laser technical noise correlations, as measured with auxiliary sensors; therefore, this band is excluded from analysis. A small number of frequency bins near cavity resonances are also excluded at higher frequencies on the basis of exhibiting an excess coherence with laser noise monitors, resulting in a total of $N=1728$ science-quality frequency bins up to 20~MHz at 9.92~kHz resolution. Our full suite of data quality tests are detailed in \S8 of Ref.~\cite{chou2017}. Several enhancements of these tests are summarized in the \textit{Supplemental Material}.

The measured complex CSD is consistent with no correlated displacement noise over the 1.1~MHz to 20~MHz band. This is seen in the final result (Fig.~\ref{fig:main_Result_v2}, bottom panels), where we present the data rebinned to 496~kHz resolution for easier visual interpretation of signal-to-noise, utilizing the statistical independence of the 9.92~kHz bins (each observing run is also found to be consistent with this null result). Band-averaged from 1.1~MHz to 20~MHz, the real and imaginary components of the final CSD are $(0.01 \pm 0.01)\,t_P$ and $(-0.01 \pm 0.01)\,t_P$, respectively, in units of strain noise power spectral density. The error bars represent the combined $1\sigma$ statistical uncertainty and 10\% systematic calibration uncertainty~\cite{chou2017}. The real and imaginary CSD components are considered separately because both conventional RF backgrounds and novel fields (e.g., axion-like couplings~\cite{Grote:2019uvn}) will generally couple to the interferometers in-phase. In this case, correlated displacement noise will manifest entirely in the real quadrature, halving the number of sensitive degrees of freedom and thus the measurement variance. Correlations arising from holographic quantum geometry~\cite{Hogan:2016}, on the other hand, depend on the antenna response to the nonlocally fluctuating background and can manifest in both quadratures. Within the 1.1~MHz to 20~MHz band, we place an average limit of $3.72\times 10^{-41} \,\mathrm{m}^2 /\mathrm{Hz} \,\approx\, 0.46\,t_P L^2$ on the magnitude of stationary sources with bandwidth $\Delta f > 9.92\;\mathrm{kHz}$. For a white broadband stationary noise spanning the entire band, this limit improves to $1.03\times 10^{-42} \,\mathrm{m}^2 /\mathrm{Hz}$.

%--------------------------------------------
\paragraph{\textbf{Model testing.}}
We use this measurement to test a semiclassical model of instrument response to conjectured coherent, spacelike rotational fluctuations arising from quantum geometry. We adopt the prediction as in Ref.~\cite{Hogan:2016}, but recalculated for the as-built instrument to account for slight differences in optical distances. The resulting model spectrum is shown in Fig.~\ref{fig:main_Result_v2} (purple curves, all panels). Its magnitude scales with a normalization parameter, nominally $\eta \approx t_P$, which conveys a spectral density of information: along any timelike world line, $\eta$ is, heuristically, the unit time for each independent quantum ``bit'' to be defined. Its inverse regulates the scaling of directional uncertainties in the local inertial frames of causal diamonds and the entropies associated with the boundaries~\cite{Hogan:2016}. We constrain such rotational uncertainties using cross-interferometer CSD data from 1.1~MHz to 20~MHz to place an upper limit on $\eta$.

To obtain this limit, we perform a likelihood-ratio test. For two shot-noise-dominated signals, the noise in each quadrature of the CSD is independent and Gaussian-distributed. Using this fact, the likelihood ratio of a given model normalization to the $\eta = 0$ (classical space-time) case assumes a simple form,
\begin{equation}
    \Lambda(\eta) \equiv \exp\left(-\frac{1}{2}\left[\chi^2(\eta) - \chi^2(0)\right]\right)\;,
\end{equation}
where
\begin{eqnarray}
    \label{eq:chisquare}
    \chi^2(\eta) &=& \sum_{k=1}^N \Bigg[ \frac{\left[ \operatorname{Re}\,D(f_k) - \operatorname{Re}\,M(f_k, \eta) \right]^2}{\operatorname{Var}\left[ \operatorname{Re}\,D(f_k)\right]} \nonumber\\
    && \;\,\,\,\, +\;\, \frac{\left[ \operatorname{Im}\,D(f_k) - \operatorname{Im}\,M(f_k, \eta) \right]^2}{\operatorname{Var}\left[ \operatorname{Im}\,D(f_k)\right]} \Bigg]
\end{eqnarray}
is the $\chi^2$ statistic of the complex spectral model. In Eq.~\ref{eq:chisquare}, $D$ refers to the CSD data (Fig.~\ref{fig:main_Result_v2}, red data points) and $M$ to the model prediction (the nominal case, $\eta = t_P$, is shown in purple curves). By Wilks' theorem, the difference in $\chi^2$ values, $\chi^2(\eta) - \chi^2(0)$, is asymptotically \mbox{$\chi^2$-distributed} with a single degree of freedom. This allows a direct estimation of the statistical significance. At 95\% confidence we constrain the normalization to $\eta < 0.25\,t_P$ for rotations around one axis, corresponding to a model with information content exceeding the conjectured bound from holographic entropy.

%--------------------------------------------
\paragraph{\textbf{Conclusions.}}
Our results demonstrate a new frontier in precision measurement. For the first time, high frequency differential lengths ($\delta l$) in a pair of large interferometers are used to measure a rotational degree of freedom. In coupling rotations of the system to the differential arm length, the bent-arm interferometers (see Fig.~\ref{fig:layout}) sense an observable that is physically distinct from both the previous Holometer instrument and gravitational wave detectors. For stationary sources with bandwidth $\Delta f > 10\;\mathrm{kHz}$, the sensitivity to rotation-induced strain surpasses $\mathrm{CSD}_{\delta h} < t_P / 2$ across a broad band from 1.1~MHz to 20~MHz. By averaging down the dominant, incoherent sources of interferometer noise over $3.9 \times 10^{10}$ independent spectral measurements per 9.92~kHz bin, we attain a sensitivity to coherent displacement noise power five orders of magnitude below the quantum sensing limit of either interferometer individually. Most importantly, we demonstrate with this technique that high-frequency environmental mitigation is possible (see the \textit{Supplemental Material} and Ref.~\cite{chou2017}), enabling experimental sensitivities well beyond the threshold needed to test phenomenological models of holographic quantum spacetimes~\cite{Hogan:2016} and couplings of axion-like dark matter fields~\cite{Grote:2019uvn}. The use of quantum optical techniques may enhance the sensitivity even further~\cite{2013PhRvL.110u3601R,Pradyumna_2020}.

The data presented here are consistent with a classical spacetime. They constrain a specific kind of departure from classicality---spacelike coherent, rotational quantum fluctuations about a single axis~\cite{Hogan:2016}---which would not have been detected before. Models of quantum gravity based on locally quantized fields on classical backgrounds predict no detectable effect in this type of measurement. Even so, searches for nonstandard, nonlocal quantum-geometrical effects are well motivated. Frameworks built on classical locality have well-known theoretical difficulties,  such as accounting for information flow in the context of black hole horizons~\cite{Hooft:2016itl,Hooft2018,Banks:2018aed,Giddings:2019vvj} and the small value of the cosmological constant~\cite{Weinberg:1988cp,*Weinberg:2008zzc,CohenKaplanNelson1999,*cohen2021gravitational,Banks:2018jqo,Hogan:2020aow}. Radical proposed modifications that address these issues, such as scale-invariant coherent geometrical states on causal diamonds, holographic nonlocal correlations, or geometrical ``squeezing''~\cite{Verlinde2019,Verlinde:2019ade,Parikh:2020nrd,*Parikh:2020kfh}, often produce large fluctuations on the  displacement scale probed here, $\mathrm{CSD}_{\delta h}\sim\hspace{.07em} t_P$. Our results significantly constrain such nonstandard options with a precision far exceeding limits currently achieved by other techniques, such as optical clocks~\cite{Campbell90,PhysRevLett.124.241301}.

Even though our measurement exceeds Planck sensitivity, it does not exclude all theories with large, nonlocally coherent holographic correlations. All light paths in our interferometers lie in a single plane. This configuration does not test models with Planck-scale uncertainty that entangles all three spatial directions; for example, it cannot coherently compare rotational states in orthogonal directions that may be associated with incompatible observables. Such models, if experimentally confirmed at Planck spectral density, may explain apparent anomalous symmetries in the cosmic microwave background\footnotemark~\cite{PhysRevD.99.063531,Hogan_2020,Hagimoto_2020}. A future experiment~\cite{grote2020} is planned in a general 3D geometrical configuration to probe this class of theories. A quantitative assessment of new experimental designs thus motivated, as well as a thorough theoretical interpretation of our current result, will follow in future work.\newpage

\footnotetext[\value{footnote}]{For example, angular correlations of CMB temperature anisotropy, the only measurement to date of quantum correlations on the scale of a causal horizon, appear to exactly vanish at an angular separation of 90 degrees~\cite{Hogan_2020,Hagimoto_2020}.}

%--------------------------------------------
\paragraph{\textbf{Acknowledgments.}}
This work was supported by Fermi National Accelerator Laboratory (Fermilab), a U.\,S. Department of Energy, Office of Science, HEP User Facility, managed by Fermi Research Alliance, LLC, acting under Contract No.~DE-AC02-07CH11359. We are grateful for support from the John Templeton Foundation, the University of Chicago / Fermilab Strategic Collaborative Initiatives program, and the Fermilab Laboratory Directed Research and Development program. J.R. was partially supported by the Visiting Scholars Award Program of the Universities Research Association (Award No.~18-S-20). O.K. was partially supported by the Basic Science Research Program (Grant No.~NRF-2016R1D1A1B03934333) of the National Research Foundation of Korea funded by the Ministry of Education. The Holometer team gratefully acknowledges the extensive support and contributions of Gregory L. Brown, Andrea Bryant, Erin Glynn, Raymond H. Lewis, Arlo Marquez-Grap, Jeronimo Martinez, Matthew Quinn, James E. Ranson, Eleanor Rath, George Ressinger, and Michael Shemanske in the construction and operation of the apparatus. We also thank Rana Adhikari and Hartmut Grote for insightful comments during the data analysis and interpretation.

%--------------------------------------------
\bibliography{holometer}
\cleardoublepage

%--------------------------------------------
\section{Supplemental Material}

%--------------------------------------------
\begin{figure*}[t]
  \centering
  \includegraphics[width=.692\textwidth,trim=0.in 0.13in 0.in 0.05in]{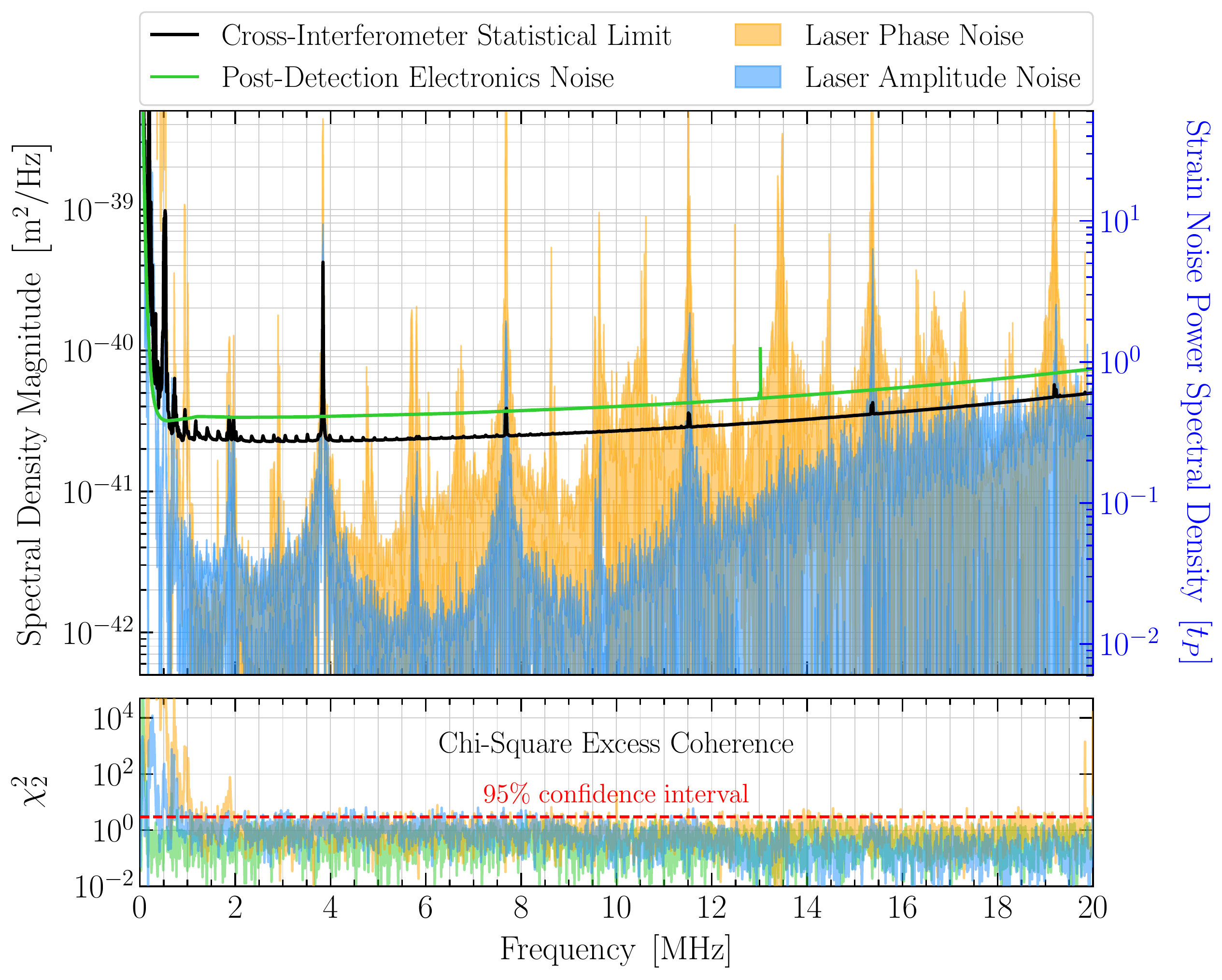}
  \caption{Measured limits on environmental sources of correlated noise. The curves show the exclusions of input laser phase noise (orange), input laser amplitude noise (blue), and post-detection electronics noise (green). For reference, the black curve shows the statistical sensitivity of the two-interferometer cross-spectrum. The electronics noise curve is from a 745~hour ex-situ measurement, in contrast with the in-situ sensitivity of the 1098~hour science data set. The reduced $\chi^2$ statistic for $k=2$ degrees of freedom, shown in the bottom panel, indicates the level of statistical certainty of coherent noise detection. Values~$>3$ are indicative of significance at the 95\% confidence level.}
  \label{fig:systematic_backgrounds}
\end{figure*}

\vspace{14pt}\subsection{Limits on Environmental Noise Couplings} % spacing adjustment

We have carried out the same set of tests for systematic noise biases as outlined in \S6.4 of Ref.~\cite{chou2017} for the first-generation Holometer experiment, with some additions and extensions as described here. The results of these tests show there would be insignificant impact on our science results from environmental RF noise. The measured limits on environmental noise couplings are shown in Fig.~\ref{fig:systematic_backgrounds}.\vspace{-16pt} % spacing adjustment

\subsubsection{Input Laser Noise}
In the original straight-arm configuration, the search for an exotic signal was carried out at frequencies below 10~MHz. Consequently, characterization efforts for correlated laser noise targeted the band below 10~MHz. For the rotational configuration, to account for the possibility of a signal extending to 20~MHz, the measured limits on correlated laser noise must be extended comensurately.

\paragraph{Laser Phase Noise.}
Measurements of the interferometer transfer function of laser phase noise to DARM were carried out as described in \S6.4.2 of Ref.~\cite{chou2017}, but with the measurement bandwidth extended to 20~MHz. The measurements were made on several nonconsecutive days and the results compared for consistency. Above 10~MHz, the phase of the transfer function was found to vary from day to day. This is likely caused by alignment-dependent variation of the couplings of higher-order spatial modes (HOMs) to the antisymmetric-port (ASP) photodetectors (PDs), as occurs upon each re-tuning of the instrument. Because the interferometers do not have output mode cleaner cavities, HOMs are not attenuated prior to detection. HOM noise sidebands beat with the fundamental-mode carrier to produce amplitude modulations of the detected power. To obtain a worst-case limit shown in Fig.~\ref{fig:systematic_backgrounds} (orange curve), we have taken the maximum coupling amplitude among all measurements for each frequency bin individually.

\paragraph{Laser Amplitude Noise.}
Measurements of the interferometer transfer function of laser amplitude noise to DARM were carried out as described in \S6.4.1 of Ref.~\cite{chou2017}, but with the measurement bandwidth extended to 20~MHz. They were found to be consistent over several days. However, because this measurement passively relies on the amplitude noise of the input laser, the coherence is poor above 10~MHz, where deviations in the phase noise case were most strongly observed. Fig.~\ref{fig:systematic_backgrounds} (blue curve) shows the amplitude noise coupling obtained from the transfer function measurement average.

\subsubsection{Ambient Radiofrequency Noise}
Ambient RF coupling to the interferometer signal takes two possible forms. The first is direct coupling to the post-detection electronics. Tests of this form of contamination (see \S6.4.3 of \cite{chou2017}) were extended to include a new white noise test at full ASP PD photocurrent. The second is excitation of the actual DARM signal, either by coupling to physical motion---e.g., exciting the piezo-electric transducer (PZT) actuators of the end-mirrors---or by an optically active effect in the optical substrate or reflective coatings, coupling electric or magnetic fields to path length. The actuations of the PZTs were tested by coupling electrical signals, either from the ambient RF environment or capacitively, to the PZT signal cables directly. These effects have not been newly tested since the first configuration. However, the addition of a \mbox{$45^\circ$-incidence} bend mirror in one arm of each interferometer introduces potential new modes of coupling directly to DARM via its sensitivity to polarization. Tests for polarization-dependent effects are detailed below.

\paragraph{Post-Detection Electronics Noise.}
As in the previous configuration, the most sensitive test of correlated post-detection additive electronic noise is a run with the photodiodes, amplifiers, and analog-to-digital converter (ADC) systems operating with the laser light off. Because the system under normal operation is highly shot-noise-limited, a relatively short 10~hour dark run serves to test the possibility of additive noise coupling to both detector signal chains and causing an apparent correlated signal. This test shows no evidence of this source of correlated signal. However, the dark test leaves open the possibility that high-noise, high-current operating conditions nonlinearly impart correlations not apparent under dark conditions. Large noise on an ADC having digitization issues, an amplifier or PD with saturation problems, or amplifier nonlinearity such as slew rate limitations could couple the detection system to external signals differently during data runs than during dark noise tests.

For this reason, the system was newly run with a separate halogen light bulb focused on each of the ASP PDs so that the photocurrent was the same as during science-mode operation. The light bulbs were optically coupled to the photodiodes with fast lenses and run with the lowest possible light bulb currents to avoid UV illumination on the photodiodes during the test. As during the science data runs, the photocurrent to shot noise ratio was monitored to verify the noise purity. The disadvantage over the dark noise test is that bright noise data must be collected for as long as the science data integration to reach the needed sensitivity to spurious correlation. We obtained 745~hours of data in this configuration, which is less than the total integration time for the science-mode configuration. Hence the statistics-limited electronics noise coupling, shown in Fig.~\ref{fig:systematic_backgrounds} (green curve), sets a slightly worse limit than the sensitivity of the science data itself (black curve). However, no correlation was found in these data over the range from 1.1~MHz to 20~MHz; this limit could be improved with a longer integration time.

\paragraph{Polarization Noise.}
New polarization-dependent couplings are introduced due to the $45^\circ$ angle of incidence of the beam at the bend mirrors. Both bend mirrors are coated for $p$-polarization. However, any small rotation of the polarization axis introduces some $s$-polarization at the bend mirror surface, modulating the penetration depth into the dielectric coating layers. Through this mechanism, ambient magnetic fields rotating the polarization of the light could modulate both the amplitude and phase of the reflected laser field. Such effects are negligible at the power recycling mirror and end mirrors, which have near-normal angles of incidence, and cancel to first order at the \mbox{$45^\circ$-incidence} beamsplitter, due to the double-passing of the light.

To check for such a possibility, a series of RF electromagnetic tests were conducted on the bend mirrors while the interferometers were running. A set of Helmholz drive coils were placed around both 10 inch stainless steel corner cubes containing the bend mirrors and energized with a power amplifier operating to 15~MHz. The coils were placed so that they were aligned along three perpendicular axes. A pickup coil placed between the cubes measured both the environmental RF and the drive coil signal. A comprehensive set of measurements were first made relative to ambient RF fields. Then, during the drive tests, a second pickup antenna was used as a witness sensor to maximize the sensitivity of the transfer function upper limit between the magnetic field at the pickup coil and DARM, similar to the method in Eq.~25 of Ref.~\cite{chou2017}. The pickup coil and witness signals were digitized at 50~MHz together with the interferometer signals and cross-correlated. Several hours of RF noise with magnetic field amplitudes 1000 times ambient was measured, using swept RF signals, broadband noise, and several discrete lines. However, the measurement was found to be limited by direct couplings of the detection electronics to the strong RF field being generated, as the measured transfer function was identical even when the interferometers were not locked. This indicates that the strongest couplings of the apparatus to ambient electromagnetic fields are not optical but electronic in nature. Thus, if ambient electromagnetic fields of sufficient strength to induce an optical effect were present, they would have also been detected as correlated post-detection electronic noise.

%--------------------------------------------
\begin{figure*}[t]
  \centering
  \includegraphics[width=.620\textwidth]{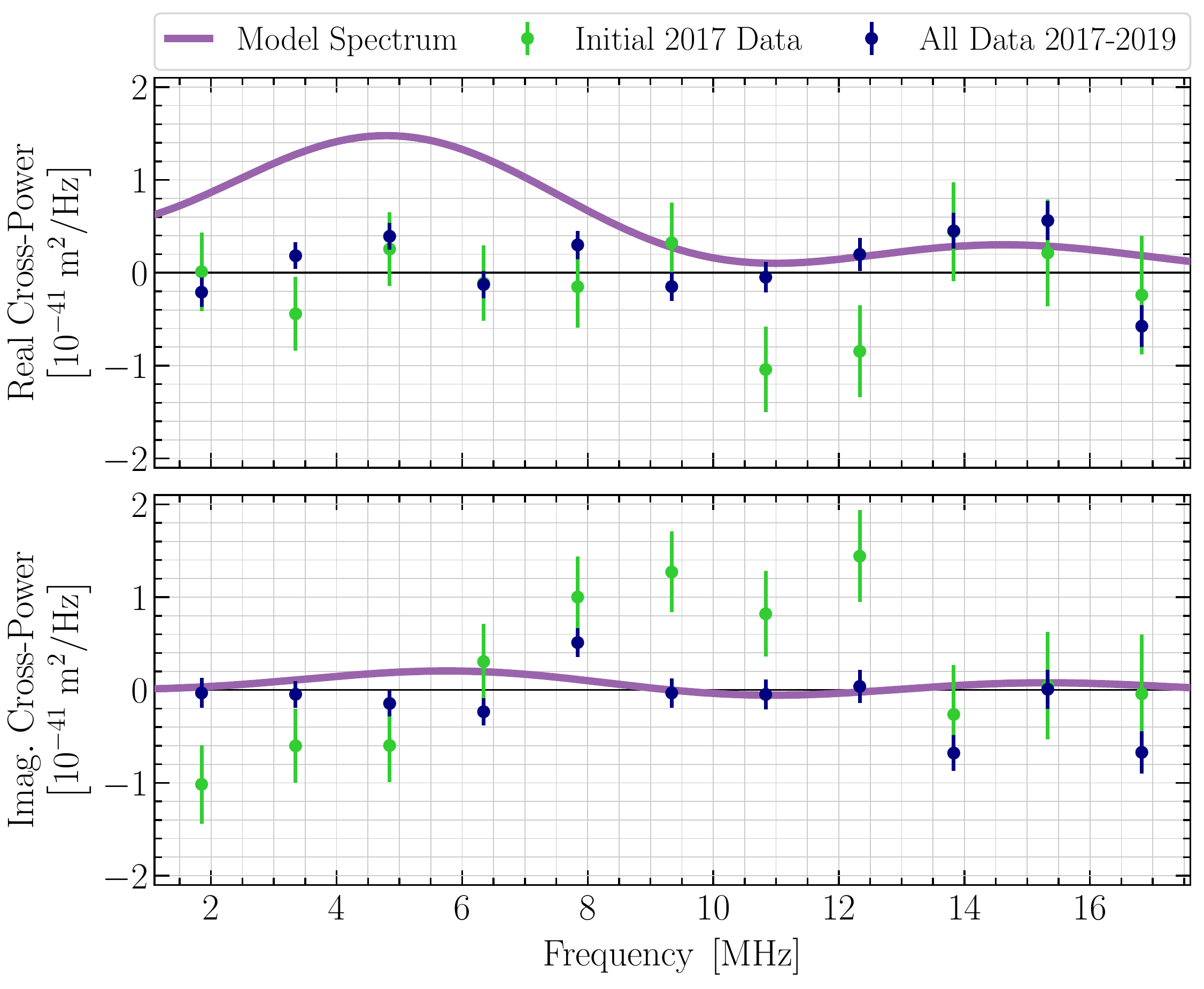}
  \caption{Interferometer CSD from Spring 2017 data (125 hours; green data points) vs. the full data set (1098 hours; blue data points). The data are shown for both real and imaginary components at 1.5~MHz resolution. For reference, all panels are overlaid with a semiclassical model spectrum of quantum geometrical fluctuations (purple curves) reproduced from Ref.~\cite{Hogan:2016}, but adjusted to account for the actual dimensions of the as-built instrument.}
  \label{fig:comparison_to_spring2017}
\end{figure*}

\subsection*{Chronology of Observing Runs}

The measurements reported here were taken over a two-year period, from April~2017 to August~2019, in five separate data runs. Almost all data were taken at night for more stable optical cavity lock, as the reconfigured bent-arm instrument had less insulation and could not maintain long-term alignment when direct sunlight caused varying thermal expansion in the vacuum tubes.

In Spring 2017, the first 125~hours of data were collected (run~1), passing all systematics tests established in the first-generation Holometer experiment prior to the reconfiguration~\cite{chou2017}. The data showed an apparent broadband correlation from 6.5~MHz to 13~MHz purely in the {\it imaginary} part of the cross spectrum, as shown in Fig.~\ref{fig:comparison_to_spring2017} (green data points). Since this feature was entirely different from the predicted model, its statistical significance was difficult to assess post facto in a model-independent or nonparametric manner.

To further investigate this effect, we collected two more sets of data throughout Summer and Fall 2017: one for 113~hours of additional science data in the nominal configuration (run~2), and another for 141~hours of data in an ``inverted fringe'' configuration (run~3). In the latter, one of the control systems was operated with a sign inversion of the DARM offset and feedback gain, so that one interferometer was locked on the opposite side of the dark fringe (IFO~1 inverted for 100~hours, IFO~2 for 41~hours; see Figs.~\ref{fig:layout} and \ref{fig:main_Result_v2}). This has the effect of inverting the sign of the optical response to DARM perturbations, as well as to other sources of phase noise, creating a phase inversion in the interferometer cross spectrum. The response to amplitude noise sources, including post-detection additive electronics noise, is unchanged. This technique is thus a diagnostic tool for determining the coupling mechanism of potential background cross-correlations. Our tests did not clearly reproduce the Spring 2017 signal, however, and further data acquisition was precluded by the rapidly decaying power levels of the two Mephisto 2~W Nd:YAG lasers (both after over 30,000 hours of use).

In Spring 2018, the lasers were refurbished with replacement pump diodes by Coherent, Inc. and reinstalled into the interferometers. This restored the laser power to levels even higher than before. In Summer 2018, only a limited set of science data amounting to 64~hours was collected (run~4), due to interferometer controls difficulties. This had insufficient statistical power to independently test the first data set. Additional systematic tests were conducted throughout the remainder of 2018, in parallel with efforts to retune the control systems for higher operational stability. From Spring through Summer 2019, a final, extended science run was conducted with reoptimized control systems, yielding 654~hours of additional data (run~5). With the combined statistical power of all data sets since Spring 2017, we obtained a conclusive rejection of the apparent effect in the first data run.

\end{document}